# AI-Enhanced High-Density NIRS Patch for Real-Time Brain Layer Oxygenation Monitoring in Neurological Emergencies

Minsu Ji[1], Jihoon Kang[2], Seongkwon Yu[1], Jaemyoung Kim[3], Bumjun Koh[1], Jimin Lee[1], Guil Jeong[1], Jongkwan choi[3], Chang-Ho Yun[2][†], and Hyeonmin Bae[1][†]*

[1]School of Electrical Engineering, Korea Advanced Institute of Science and Technology, Daejeon, Republic of Korea

[2]Department of Neurology, Seoul National University Bundang Hospital and Seoul National University College of Medicine, Seongnam, Gyeonggi-do, Republic of Korea

[3]Department of Research and Development, Optics Brain Electronics Laboratory, OBELAB Inc, Seoul, Republic of Korea

[†]The two authors share the senior authorship.

**\*Corresponding author**

E-mail: hmbae@kaist.ac.kr

## Abstract

Photon scattering has traditionally limited the ability of near-infrared spectroscopy (NIRS) to extract accurate, layer-specific information from the brain. This limitation restricts its clinical utility for precise neurological monitoring. To address this, we introduce an AI-driven, high-density NIRS system optimized to provide real-time, layer-specific oxygenation data from the brain cortex, specifically targeting acute neuro-emergencies. Our system integrates high-density NIRS reflectance data with a neural network trained on MRI-based synthetic datasets. This approach achieves robust cortical oxygenation accuracy across diverse anatomical variations. In simulations, our AI-assisted NIRS demonstrated a strong correlation ($R^2=0.913$) with actual cortical oxygenation, markedly outperforming conventional methods ($R^2=0.469$). Furthermore, biomimetic phantom experiments confirmed its superior anatomical reliability ($R^2=0.986$) compared to standard commercial devices ($R^2=0.823$). In clinical validation with healthy subjects and ischemic stroke patients, the system distinguished between the two groups with an AUC of 0.943. This highlights its potential as an accessible, high-accuracy diagnostic tool for emergency and point-of-care settings. These results underscore the system's capability to advance neuro-monitoring precision through AI, enabling timely, data-driven decisions in critical care environments.

Brain monitoring has long been one of the most important tasks in the medical field, as neurological disorders are the second leading cause of death and the leading cause of disability in the world[1-4]. In particular, neurological emergencies that require treatment within a certain window of time, including stroke, intracerebral hemorrhage, and cerebral edema, have resulted in the development of an accurate diagnostic device capable of fast measurement and continuous monitoring[5-7]. Despite recent advances in brain monitoring devices, magnetic resonance imaging (MRI) and computerized tomography (CT) are two main imaging modalities used for the diagnosis of neurological diseases. However, limited accessibility and low portability make such techniques inappropriate for point-of-care (POC) use or the continuous monitoring of patients who require intensive care. Such limitations are especially critical for individuals with time-sensitive diseases, such as acute ischemic stroke, in which delayed treatment can cause permanent damage to the brain or even death[5]. In addition, the benefits of these expensive devices are limited in countries with large populations and inadequate medical facilities. Other brain monitoring modalities, such as microwave, ultrasound, and electroencephalogram (EEG), are rarely used due to their limited spatial resolution[8-11].

Continuous wave near-infrared spectroscopy (CW-NIRS) is a noninvasive, low-cost, and easy-to-use technology that is used to measure the concentration of oxy/deoxy hemoglobin in the tissue and has the potential to be used to measure other critical biomarkers, such as the water fraction and concentration of cytochrome C oxidase[12-15]. Commercially available NIRS devices that are approved for use by the FDA (U.S. Food and Drug Administration) employ the scheme of spatially resolved spectroscopy (SRS), which is based on the assumption that the head anatomy can be modeled by a homogeneous semi-infinite slab, and this approach can be used to analytically extract the fraction of oxyhemoglobin concentration ($C_{oxy-Hb}$) in the total hemoglobin ($C_{oxy-Hb}$ +

$C_{deoxy\text{-}Hb}$) [13,16,17]; this metric is referred to as cerebral oxygenation (rSO2). However, these SRS devices show high susceptibility to structural variation in the head since SRS technology is generated based on the assumption that the brain is a homogeneous medium; thus, this approach fails to extract the true rSO2 of the cortex. In addition, although the light scattering coefficient of the tissue has interpersonal differences, SRS devices rely on the procrustean scattering coefficient based on empirical findings[12,18,19]. As such, the rSO2 value obtained from conventional SRS devices cannot be used to classify healthy subjects and patients with brain disease[20-25]. Several studies have tried to use diffuse optical tomography (DOT) to image brain regions, but the scattering of the traveling photon in the turbid brain medium severely limits spatial resolution[26-28].

This article proposes a novel neural network-assisted high-density NIRS system that can accurately extract information about brain cortex layers, overcoming the limitations of current brain monitoring devices (Fig. 1). The system consists of a pair of flexible high-density NIRS patches, a control board, and a tablet PC that is used to operate the brain layer analyzing network. The flexible patches, including components for sensing and signal processing, collect the multiwavelength reflectance of the brain layer. Then, the neural network extracts the brain layer information by using learning features trained with magnetic resonance (MR) images and photon simulations. To validate the proposed system, a biomimetic multilayer phantom consisting of skin, skull, cerebrospinal fluid (CSF), and brain layers was fabricated based on MR images and optical human boundary conditions. Multilayer phantom experiments were performed to measure rSO2 while varying the surface property, thickness of the CSF layer, and oxygen concentration of blood in the brain layer. In addition, to investigate the clinical effectiveness of the proposed system, the rSO2 in healthy subjects and ischemic stroke patients were compared.

# Results

## *Brain layer analyzing network*

The propagation of the NIR light injected into the biological tissue is determined by the optical properties of the medium. The absorption coefficient ($\mu_a$) and the reduced scattering coefficient ($\mu_s$'), which are quantified optical properties of a tissue, depend on the concentration of biological components, including oxygenated hemoglobin, deoxygenated hemoglobin, water, lipids, cytochrome C oxidase, melanin, and bone mineral, especially in the brain region[13,18]. The multiwavelength NIR light attenuation provides information on each constituent. However, it is very challenging to quantify the concentration of each chromophore in the brain, as the detected NIR light traveled through multiple complex layers, including skin, skull, CSF, gray matter and white matter, with different $\mu_a$ and $\mu_s$' values. We employed a supervised learning technique, which is highly effective in predicting the target label under complex input dimensions, to analyze the brain layer information from the detected NIR light distribution. For this purpose, a dataset synthesis platform using MR images and photon simulation was created for training.

Figure 2a shows the flowchart of the dataset synthesis platform. First, the anatomy from the MRI slices was segmented into five layers with different optical properties, including skin, skull, CSF, gray matter, and white matter[29]. In addition, the optical properties of each layer were mapped to the anatomy. In particular, each segmented layer was assumed to have a random concentration of biological components within the plausible range of these components in a human brain, and corresponding optical properties were mapped to each layer (Fig. 2b). The absorption coefficient of the biological tissue can be represented by

$$\mu_a = B\, rSO2\, \mu_{a, HbO2} + B(1 - rSO2)\mu_{a, HbR} + W\mu_{water} + F\mu_{a,fat} + M\mu_{a,melanin} +$$

$$C\mu_{a,oxCCO} + N\mu_{a,mineral},$$

where *rSO2* is cerebral oxygenation, *B* is the average blood volume fraction, *W* is the water fraction, *F* is the lipid fraction, *M* is the melanin fraction, *C* is the concentration of cytochrome C oxidase, *N* is the bone mineral fraction, and $\mu_{a,X}$ denotes the absorption coefficient of the *X* component[18]. The range of the factors was retrieved from previous works (Supplementary Table 1)[18]. The mapped oxygenation level in the gray matter layer, referred to as rSO2, was used as the label of the brain layer that was used to analyze the neural network. For light scattering, the reduced scattering coefficient spectrum was modeled as the weighted sum of Rayleigh and Mie scattering, where Rayleigh scattering refers to scattering by particles much smaller than the wavelength of light and Mie scattering refers to scattering by a particle of any size. The reduced scattering coefficient can be expressed by

$$\mu_s' = a\left(f_{Ray}\left(\frac{\lambda}{500\ (nm)}\right)^{-4} + (1-f_{Ray})\left(\frac{\lambda}{500\ (nm)}\right)^{-b_{Mie}}\right)$$

where *a* is $\mu_s'(\lambda=500\ nm)$, $f_{Ray}$ is the Rayleigh scattering factor, and $b_{Mie}$ is the Mie scattering factor[18]. The range of the factors was described in previous works (Supplementary Table 2).

The next step of the dataset synthesis platform involves rendering the segmented MRI slices into a 3D anatomy to simulate the patch's positional dependence, including the sensor array and light emitters, as illustrated in Fig. 2a. Light scattering leads to differences in the information obtained from detectors placed close to the light source and those situated at a distance. The former provides insight into the structure of shallow layers, while the latter reflects a mixture of information from shallow and deep layers[13]. To enhance the analysis of shallow layer structures, a pair of laser sources is positioned at either end of the sensor array, providing additional data from multiple regions. The optimization of the sensor array size for maximizing the performance of the neural network is explained in a later section. Curve fitting was used to place the flexible patches onto the highly curved surface of the anatomy. The nasion, which is the middle point of the nasofrontal suture, was used as the reference point to generalize the position of the patch in an arbitrary subject (see Methods for details).

In the last step of the dataset synthesis platform, Monte-Carlo photon simulations were conducted at four distinct wavelengths (725 nm, 780 nm, 850 nm, 940 nm) to produce 2D optical density (OD) images, as shown in Fig. 2a[30]. The OD from the i[th] detector is represented by

$$OD_i(\lambda) = -ln\left(\frac{I_i(\lambda)}{I_0(\lambda)}\right),$$

where $\lambda$ is the wavelength of the light, $I_0(\lambda)$ is the incident light and $I(\lambda)$ is detected light from the i[th] photodetector. Estimating the incident light intensity $I_0$ presents a challenge due to variations in both the direction of emitted light and the reflection coefficient of skin. To mitigate the impact of the unknown $I_0$, the intensity of the 2D OD image is normalized by that of the first photodetector $OD_1$. The 780 nm and 850 nm wavelengths are widely used for the evaluation of hemoglobin concentration and rSO2[13,31,32]. The 725 nm wavelength was used to increase the accuracy of the rSO2, and the 940 nm wavelength was used to increase the sensitivity to water[33]. Finally, the dataset synthesis platform acquires eight OD images using two emitters that have four wavelengths each.

We generated the dataset with the rSO2 label of the gray matter using the dataset synthesis platform to train the brain analyzing network. One thousand random optical characteristics were mapped to the anatomy originating from real MR images of 600 subjects to obtain a total of 600,000 distinctive data. Figure 2c shows the architecture of the neural network. MobileNet V2 was used to minimize the inference time and the size of the network[34]. The model utilized the concatenation of the eight optical density maps obtained from the pair of four wavelengths emitters as the input, and the rSO2 of the gray matter was the designated output label of the network. The loss of the network is the root mean square error (RMSE) between the true rSO2 of the gray matter and the prediction of the network, which can be expressed by

$$Loss = \sqrt{\frac{1}{N}\sum(\widehat{rSO2}_{brain} - rSO2_{brain})^2},$$

where $rSO2_{brain}$ is the actual rSO2 value of the gray matter, $\widehat{rSO2}_{brain}$ is the prediction, and $N$ is the batch size of the training.

Figure 3 shows the design procedure for the sensor array, in which the performance of the brain analyzing network was maximized. First, we determined the size and spacing of the photodetectors. A dense array provides high spatial resolution, but the upper limit of the density is predominantly influenced by the resolution of the ground-truth MR images used for the photon simulation. In an ideal scenario, all photodetectors would have equal intersection areas, but due to the limited spatial resolution of the MR image, the simulated light intensity experiences quantization errors, as illustrated in Fig. 3a. This quantization error is especially critical for small photodetectors. The simulated light intensity $\hat{I}_{sim}$ of the photodetector is determined by

$$\hat{I}_{sim} = E_e\hat{A} = E_e(A + A_{qe}) = E_eA(1 + r_{qe}) = I_{sim}(1 + r_{qe})$$

where $E_e$ is irradiance flux density (W/m²), $\hat{A}$ is intersection area between the photodetector and the surface of the MR image, $A$ is the ideal intersection area of the photodetector, $A_{qe}$ is the quantization error in the intersection area of the photon detector, $r_{qe}$ is the quantization error ratio, $A_{qe}/A$, and $I_{sim}$ is the error-free light intensity. The variance in the intersection areas among the photodetectors finally results in the quantization error in the simulated $OD$ spatial distribution, which can be expressed by

$$\widehat{OD} = OD + OD_{qe},$$

where $\widehat{OD}$ is the photon simulation result using a resolution-limited MR image, $OD$ is the error-free underlying optical density, and $OD_{qe}$ is the quantization error in the optical density due to the limited resolution. Considering that the $\sigma(OD_{qe})$ can be estimated as the $\sigma(r_{qe})$, $\sigma(r_{qe})$ is computed through simulation results that assess the variance in the intersection area of the photoreactors in response to changes in the size of the photodetector (see Supplementary Note 1 for details). Figure 3b illustrates the $\sigma(OD_{qe})$ relative to the size of the photodetector. The chosen dimension of each photodetector is 4 mm, which is sufficiently large to resist the effects of pixilation in the MR image.

Second, the performance of the network was examined while the number of rows and columns of the sensor array were changed to investigate the optimal optical arrangement, as shown in Fig. 3c. Since the network did not converge when the number of rows was less than 8 in the training phase, we increased the number of rows from 8. Figure 3d represents the accuracy of rSO2 extraction with respect to the number of rows and columns of the sensor arrangement. A 12 by 5 sensor array size was determined to be the optimal choice for maximizing network performance, achieving an RMSE of 5% while minimizing the number of photodetectors.

To evaluate the performance of the proposed brain analyzing network and compare it with conventional SRS techniques, an additional evaluation set of 150,000 data points was generated using the dataset synthesis platform (see Supplementary Note 2 for details). The evaluation set was generated from MR images of 150 subjects that were not used for the training. Figure 3e demonstrates the rSO2 extraction performance of the proposed system and the SRS technique. The rSO2$_{proposed}$ was highly correlated with the true rSO2 of the gray matter ($R^2$=0.913), while the rSO2$_{SRS}$ was less correlated ($R^2$=0.469).

To design the hardware, we measured an error tolerance of the network. An allowed signal-to-noise ratio (SNR) specification was established by adding random Gaussian noise ($OD_n$) to the optical density (OD) images of the dataset. Figure 3f presents the performance of the network under noisy OD conditions. The network exhibited stable performance when $\sigma(OD_n)$ was less than 0.12, which can be translated to an SNR level (see Supplementary Note 3 for details).

## *High-density NIRS system*

Figure 4a demonstrates a noninvasive portable patch system for continuous real-time measurements of true rSO2 of the cortex. The system consists of a pair of patches with a high-density sensor array to monitor the left and right sides of the forehead, a control board, and a tablet PC that is used to operate the brain analyzing network. The width, length, thickness, and weight of the patch are 37.7 mm, 81.3 mm, 3.8 mm, and 12.5 g, allowing for easy placement on the human forehead. The control board supplies power to the patches and controls the laser and sensor array on the patch to obtain the optical density map. The external tablet PC (Surface pro 7) supplies power to the overall system over a USB line and translates the measured optical reflectance into rSO2$_{brain}$ by using the brain analyzing network.

A key feature of the patch system is the inclusion of a high-density sensor array while maintaining sufficient flexibility for user convenience. Figure 4b shows an exploded-view illustration of the constituent layers and components: the top silicone encapsulation layer;

electronic components including TIA array, flexible printed circuit board (fPCB), sensor array; and bottom silicone layer. A 12 by 5 photodetector array and two laser diodes are placed on an area of 52 mm by 22 mm with 4 mm spacing. As previously stated, the pixel size of the OD image was set to 4 mm. However, the actual detector size in the patch should be smaller than 4 mm for the flexibility of the patch. This discrepancy between the measurement and the simulation resulted in a random error in the OD measurement, denoted as $OD_{e,det}$. The variation in $OD_{e,det}$ was examined while the detector size was changed to investigate the accuracy degradation. Figure 4c shows $\sigma(OD_{e,det})$ for the shortest (4 mm) channel, which was the most sensitive to detector size, indicating that the error level remained below 0.04 even with detectors measuring 2 mm by 2 mm. To mitigated this error, the target SNR of the system set as 18.9 dB considering the total allowed error of the brain layer analyzing network, which was calculated in the previous section (see Supplementary Note 3 for details). The customized laser diode package includes four-wavelength emitters. Consequently, a total of 480 channels, defined as the distinct light path generated by a pair of a photodetector and a source laser, are placed on the area of 20.8 cm². The channel density of the system was 46 times greater than that of a conventional commercial device (Supplementary Table 3). Each TIA is attached to the opposite side of the photodetector to minimize the length of the high-impedance traces to suppress noise coupling. Figure 4e shows a photograph of a fabricated patch. The laser diode package is custom designed to consolidate four different emitters into a single package module, reducing the distances among the emitters to less than 1 mm (Fig. 4e). Figure 4g demonstrates the bending and twisting performance of the patch, which is crucial for the prevention of light leakage under long-term measurement. Entire photodetectors share a single variable gain amplifier (VGA) and an analog-to-digital converter (ADC) to reduce hardware complexity while minimizing performance degradation due to component mismatch. This scheme increases the flexibility of the patch by reducing the number of electrical components and decreases the overall power consumption without performance degradation.

The photon simulation results from 600 anatomies were used to determine the dynamic range of the system (see Supplementary Note 4 for details). The system is designed to have a 140 dB dynamic range so that each sensor can receive signals from two lasers placed at both ends of the patch irrespective of interpersonal variations by using dynamic control of the emission power of the laser diodes and the gain of the VGA (Supplementary Fig. 4). To extend the dynamic range, the system utilizes both the stimulated emission region and spontaneous emission region of the laser diode, which results in non-linearity in laser output power control (see Supplementary Note 4 and Supplementary Fig. 5 for details). To mitigate the impact of non-linearity in laser power control and variations in laser diodes, the OD image was constructed based solely on the relative ratios among the optical densities of the photodetectors (see Supplementary Note 5 and Supplementary Fig. 4 for details).

Figure 4g provides a block diagram of the system. The lasers controlled by the current sources (CS) on the control board emit four wavelengths of light sequentially. The light traveling through human anatomy is detected by a sensor array consisting of 60 photodetectors. TIAs with a gain of $10^7$ convert the detected photocurrent to a voltage signal. The voltage signals from all sensors are consolidated by two stages of analog multiplexers and transmitted to the control board. The multiplexed signal was amplified by a subsequent VGA and digitized by a subsequent ADC. The noise analysis was performed for the design of the system to achieve the target SNR of 18.9 dB, which ensures robust performance of the brain analyzing network (see Supplementary Note 3 and Note 6 for details). A microcontroller unit (MCU) transmits the digitized signal to the PC using USB. To receive light of a wide range of intensities, the MCU dynamically controls the gain of the VGA and adjusts the output power of the CS. Custom monitoring software on the PC performs noise filtering and generates the optical density map. The software extracts the rSO2 of the brain layer from the optical density map using the brain layer analyzing neural network. The results are plotted on the display and saved in the database using SQLite. The software based on PyQt also provides buttons to start or stop the measurement. Additionally, a differential signaling scheme is employed to suppress noise from diverse noise sources.

The system was designed to facilitate rapid and efficient measurements of oxygen saturation in supine patients. Employing a patch-type configuration, the device can be conveniently attached to patients, delivering swift oxygen saturation results within a span of 30 seconds—particularly beneficial for immediate assessments of patients within ambulance setting. To verify the system's capability to track cerebral oxygen saturation, a breath-holding test was conducted. The participant was fitted with the device and instructed to hold his breath as long as possible. The results indicated a decrease in oxygen saturation after approximately 2 minutes of breath-holding. Following the resumption of breathing, oxygen saturation exhibited an overshoot before returning to baseline levels. These findings confirm the deep learning-based oxygen saturation measurement system's efficacy in accurately monitoring human oxygen saturation dynamics.

*Biomimetic multilayer phantom experiment*

A key feature of the proposed system is the capability to extract cortex information regardless of structural and chemical variations, including variations in anatomy, cerebral blood volume linked to the concentration of hemoglobin, and CSF depth. In particular, variation in CSF depth can result not only from interpersonal variance but also from brain disorders, such as brain atrophy and edema[35].

A multilayer phantom that imitates the anatomy of a human head was designed to evaluate the intersubject tolerance of the system. Figure 5a shows the phantom consisting of the brain layer, CSF layer, and surface layer. The surface layer phantom was made of silicone and represented both the skin and skull layers. MR images were employed to make molds of the skin and skull layer phantoms (Fig. 5b). Each MR image was segmented and separated to generate 3D images of skin and skull layers[29]. The molds of the skin layer and skull layers were fabricated independently using a 3D printer. India ink and titanium dioxide were employed to set the optical properties of the phantom to cover the wide interpersonal variation of human biological tissue[36]. The four surface phantoms, based on different MR images, represent different anatomies covering diverse structures and optical properties of skin and skull layers (see Methods for details).

To control oxygenation, the brain layer of the multilayer phantom employs a liquid phantom using human blood. The recipe used for the liquid phantom was retrieved from a previous work[24]. The liquid phantom consists of human blood, SMOFlipid, phosphate-buffered saline (PBS), sodium bicarbonate buffer (SBB), yeast, and glucose, which characterize the absorption coefficient, reduced scattering coefficient, and pH of the human boundary condition. Hemoglobin in human blood enables changes in oxygenation levels in the brain layer. The oxygenation level was controlled by yeast and oxygen injection through the inlet. Additionally, the temperature of the liquid phantom was kept at 36.5°C (human body temperature) since the oxygenation of hemoglobin is sensitive to temperature[37]. The CSF layer was filled with purified water. To separate the CSF layer and the brain layer, a brain-shaped transparent silicone film with a thickness of 1 mm was fabricated using segmented MR images and used (Supplementary Fig. 8). The multilayer phantom was designed to adjust the depth of the CSF layer.

Figure 5c shows an experimental setup. To compare the performances of the proposed system and the conventional NIRS oximeter, a commercial NIRS device (O3, Masimo) was employed. The proposed system and the commercial NIRS device were attached to the left and right sides of the skin layer, respectively (Supplementary Fig. 9). A gas analyzer was used as the reference device to measure the actual SO2 of the brain layer ($SO2_{brain}$). The liquid phantom was continuously sampled by a syringe, and the sample was analyzed by a gas analyzer to measure pO2 and pH. The $SO2_{brain}$ was derived from the pO2 and pH using the hemoglobin-oxygen dissociation curve[37]. Three main variables were adjusted to simulate individual differences in the multilayer phantom (Supplementary Table 4). First, we employed four different surface phantoms, which have different structures and optical properties of the skin layer and skull layer. Second, experiments with different hemoglobin concentrations (70 µM, 100 µM, 130 µM) were performed to simulate different cerebral blood volumes (CBVs). Finally, experiments with CSF depths of 1 mm, 3 mm, and 5 mm were performed.

Figure 5d shows the results of the phantom experiments. The square of the correlation between $rSO2_{proposed}$ and $rSO2_{brain}$ was 0.986, while the square of the correlation between $rSO2_{commercial}$ and $rSO2_{brain}$ was 0.823, which indicates that the proposed system was more robust to anatomical variations than a commercial NIRS device. The results obtained through varying the hemoglobin concentration in the brain layer showed that the square of the correlation between $rSO2_{proposed}$ and $rSO2_{brain}$ was 0.987, while the square of the correlation between $rSO2_{commercial}$ and $rSO2_{brain}$ was 0.901. When the CSF depth was varied, the square of the correlation between $rSO2_{proposed}$ and $rSO2_{brain}$ was 0.990, while the square of the correlation between $rSO2_{commercial}$ and $rSO2_{brain}$ was 0.872.

The results of the multilayer experiments demonstrate that the proposed system can be used to extract brain layer information accurately regardless of anatomy, the concentration of hemoglobin, and CSF depth, unlike conventional NIRS devices based on SRS technology.

***Prospective observational study***

Ischemic stroke reduces the oxygen supply to the cerebral artery, which decreases the rSO2 in the ischemic stroke core and penumbra, which is the reversibly injured brain tissue around the core[38,39]. Therefore, the rapid speed and high portability of the proposed system could be highly effective for monitoring acute ischemic strokes, considering that the earliest diagnosis of a penumbra is essential to treat acute stroke[39]. Although some studies have attempted to diagnose ischemic stroke patients using conventional NIRS devices, the devices exhibited poor accuracy due to error from the homogeneity assumption[8,21]. To evaluate the clinical capability of the proposed system, we performed a prospective observational study on healthy subjects and ischemic stroke patients. Forty-eight healthy volunteers between the ages of 31 and 80 and without any history or evidence of brain disease participated in this study at the Seoul National University Bundang Hospital (SNUBH) from March 2021 to November 2021. Additionally, 8 patients with large vessel occlusion (LVO) were recruited at the emergency room and stroke unit of the SNUBH from September 2020 to March 2021. Only patients satisfying the following conditions were allowed to participate in the study: $T_{max}$ (time to maximum) > 6 s on perfusion-weighed imaging or lesion in diffusion-weighted imaging on the forehead region under the patches[40]. Figure 6a and b present the imaging results obtained from the patients, which indicate that the delay in $T_{max}$ imaging or the lesion in the diffusion weighted imaging were observed in the forehead region. The 1st to 6th patients were diagnosed using perfusion-weighted imaging, while the 7th and 8th patients were diagnosed using diffusion weighted imaging. The participants in both groups were assessed using the proposed system while in the supine position.

Figure 6c shows the rSO2 distributions of the healthy subjects (blue) and stroke patients (red). The rSO2 results of the left and right sides in healthy subjects are provided, whereas the results of the ipsilateral side of the lesion in the stroke patients were displayed to investigate the decrease in the oxygen level in the lesion. The rSO2 in the healthy subjects ranged from 50.1% to 61.6%. The averaged rSO2 in the healthy subjects was 52.5%, which closely aligns with cortex-specific rSO2 values reported in studies utilizing MRI and positron emission tomography (PET)[41,42]. Figure 6d depicts a box plot of the difference in rSO2 between the healthy group and the stroke patient group. The red line indicates the median of the boxplot, whereas the blue box indicates the 25th and 75th percentiles of the results. A significant difference in the rSO2 results between the two groups was observed ($p<0.001$, paired t-test). In addition, the system led to an area under the curve (AUC) of 0.943 in the receiver operating characteristic (ROC) curve used for ischemic stroke diagnosis, which emphasizes the clinical utility of the system, as shown in Figure 6e. Additionally, a significant difference between the rSO2 in stroke patients and healthy subjects aged 70 years or over was observed ($p<0.01$, Mann–Whitney U test), thereby indicating that the differences in rSO2 were caused by the decrease in oxygen levels due to ischemic strokes (Supplementary Fig. 10).

**Conclusion**

We have reported a patch-type brain layer analyzer that uses a neural network and a high-density NIRS system to selectively extract cerebral information to provide high portability, low cost, fast measurement, and continuous real-time monitoring of brain disease. The brain layer analyzing network, which learns features for the optical and structural human boundary conditions using MR images and optical properties of the biological tissue, was incorporated into the compact system, which included flexible patches with high-density sensing and processing components. The multilayer phantom experiments demonstrated the capabilities of the system in comparison to conventional NIRS devices that exhibit low accuracy due to errors caused by overestimation for individual anatomy and optical properties. Finally, we investigated the clinical utility of the proposed system by conducting a prospective observational study on healthy subjects and ischemic stroke patients.

The proposed system can potentially shorten the time period before patients are treated with time-sensitive neurological diseases, such as acute ischemic stroke, thereby preventing long-term brain damage and providing a higher chance of survival. Furthermore, the high portability and rapid measurement enabled by the system make diagnosis possible even in an ambulance. Additionally, we expect the system to contribute to the continuous monitoring of patients after an operation. For example, continuous monitoring after endovascular therapy (EVT), especially during the first 24-48 hours, is essential to prevent procedural complications, considering that 4% to 29% of post-EVT patients exhibit at least one complication, including intracranial and infarct hemorrhage[43].

This technology can be scaled to analyze other useful biomarkers, including water fraction, cytochrome C oxidase levels, and the contents of CSF, by replacing the output label of the brain analyzing neural network. In addition, for clinical application of the device, evaluations are required to measure the other sides of the head, considering that the middle central artery (MCA) territory is the most commonly affected region in neurological diseases[44]. However, because the hair in these regions of the head causes strong light absorption, further research is needed to improve optical contact[45]. Therefore, we expect that the combination of multiple biomarkers at various measurement regions could broaden the applications of this system, thereby advancing the standards for cerebral monitoring.

**Methods**

*Study design*

The core objective of this study was to validate the effectiveness and reliability of our innovative deep learning-enhanced Near-Infrared Spectroscopy (NIRS) system for diagnosing acute ischemic stroke, focusing on its adaptability, accuracy, and comparison with conventional NIRS devices. The study received approval from the Institutional Review Board of Seoul National University Hospital (B-2006-616-301, B-2101-658-302). Informed consent was obtained from all participants before the commencement of the study. A total of 37 healthy subjects (18 male, mean age = 52.3) and eight ischemic stroke patients (mean age = 72.2) were recruited; the former assessed in supine positions on both sides of the forehead region and the latter from the emergency room and stroke unit in Seoul National University Bundang Hospital. Measurements for patients were acquired pre or within 24 hours post-recanalization treatment, with the lesion side specifically being measured to analyze rSO2 levels using the proposed system. Preceding each measurement acquisition, a thorough cleaning of the forehead region was conducted with an alcohol swab to negate disturbances from foreign substances. The study comprised controlled experiments and clinical observations, using a biomimetic multilayer phantom and high-density NIRS patches respectively, to test the system's accuracy and reliability under varied anatomical and environmental conditions.

Results obtained were meticulously analyzed and compared against those from commercial NIRS devices and actual medical diagnoses, providing a comprehensive insight into the system's enhanced precision and reliability in practical, real-world scenarios. For full details, see the Supplementary Materials.

*Brain layer analyzing network*

The dataset synthesis platform was developed in MATLAB 2019b. The MR images were acquired from open MRI sources (OpenfMRI)[46]. An SPM12 toolbox was used to segment the MR images into 5 layers, that is, the skin, skull, CSF, gray matter, and white matter[29]. Images with low-quality segmentation were excluded from the dataset. The segmented MR images were then smoothed using the spatial Gaussian filter to normalize the image sizes ($\sigma = 3$). The smoothing also reduces the quantization error due to the variance in the area of intersection between the photodetector and surface of the MR anatomy in the photon simulation. A random position was simulated for the sensor array, 1–4 cm upward and 0–3 cm to the side from the nasion while considering the human facial structure. Random positioning enables the neural network to learn inputs from the sensor arrays at arbitrary positions. The MCX was employed for the light transport simulator[30]. The simulations were computed on GTX1080 and RTX2080. A total of 750,000 data points were generated from the synthesis platform, 600,000 of which were used as the training set, and 150,000 were used as the test set. An implementation of the neural network was developed on the PyTorch framework and training was performed using a batch size of 128 and a learning rate of 0.01. The RMSE between the network output and rSO2 of the brain layer served as a loss function during the training.

*Fabrication of the sensor patch and control board*

The sensor patch is based on a flexible printed circuit board (fPCB) with electronic components. The four-layer fPCB comprises photodetectors (SFH2704, OSRAM Opto Semiconductors) with a radiant sensitive area of 1.51 mm$^2$, amplifiers (LMC6035, Texas Instruments), two-stages of 8:1 multiplexer (MAX4999, Maxim integrated), and custom laser diodes (Optowell). The light emitters at four different wavelengths (725, 780, 850, and 940 nm) were incorporated into a single package of laser diodes. The molded PDMS (Sylgard 184, Dow Corning, mixing ratio of base to curing agent of 10:1) encloses the fPCB for ambient light shielding.

The control board is implanted on the six-layer PCB, including a microprocessor (STM32F723ZE, STMicroelectronics), current sources (ADN8810, Analog devices), VGAs (AD8330, Analog devices), and ADCs (LTC2203, Linear technology). The patches and control board are connected using micro coaxial cables (KEL Corp.) (Supplementary Fig. 11).

*Monitoring software*

Custom monitoring software was developed based on Python 3.7 to develop the graphical user interface (GUI) and to perform signal processing and neural network inference. The GUI in the software was run on the PyQt5 package, which includes the transient graph display and software toggle switches that enable a USB connection and activate the measurement system. The software receives the streamed data from the control board and converts the raw data into optical density images, followed by neural network inference based on the PyTorch framework. The neural network extracts the rSO2 of the brain layer from the optical density image, and the results of the left and right patches are plotted on the display simultaneously. Furthermore, the raw data, measurement time, and comments written by the user are saved in the database file.

*Biomimetic multilayer phantom*

The surface phantoms comprising the skin and skull layers were fabricated using MR images that were not used to generate the training set for the network. The images were segmented using the SPM12 toolbox, and the skin and skull layers were separated from the anatomy. The 3D printer (3D WOX, Sindoh, PLA filament) was employed to generate a mold for the skin layer. PDMS (Sylgard 184, Dow Corning) was molded with India ink (Higgens) and titanium dioxide (rutile form, Junsei Chemical) to characterize the absorption coefficient and reduced scattering coefficient (Supplementary Table 4). Additionally, a small sample was fabricated from the same material to measure the optical properties of the skin layer phantoms using Imagent (ISS Inc). The above process was repeated for the skull layer. However, the skull layer phantom was molded on the skin layer to prevent the formation of air bubbles between the two layers. The four surface phantoms with different structures, absorption coefficients, and scattering coefficients were produced from the MRI images obtained from different subjects. The brain-shaped film phantom was fabricated using 1 mm of the outer surface of the gray matter layer in the segmented MR images. The fabrication process used for the film and surface phantoms were identical, except that India ink and titanium dioxide were not used to minimize an unexpected effect due to the film phantom. All silicone phantoms were degassed under a heating or vacuum chamber to eliminate air bubbles.

A brain layer was implemented in the multilayer phantom using the liquid phantom with human blood. The recipe used for the phantom was retrieved from a previous work[24]. Eighty-nine milliliters of SMOFlipid 20% solution (Fresenius Kabi) was added to 3000 mL of phosphate-buffered saline (PBS, pH=7.4, Gibco) in a custom stainless container, characterizing a human boundary condition. We assumed that SMOFlipid could deliver scattering properties that were similar to those of Intralipid in a previous study, due to their similar compositions. The pH of the phantom stabilized to 7.4 once 30 mL of sodium bicarbonate buffer was added (SBB 8.4%, 1 mmol/ml, *Alfa Aesar*), considering that the changes in pH cause a large shift in the oxygen–Hb dissociation curve. The human erythrocyte blood bag (Hematocrit = 71%, Korean Redcross) was added, which characterized the Hb concentration pertaining to the absorption spectrum. Yeast (1.5 g, dry yeast, Jeonwon Food) and 4.5 g glucose (*SIGMA-Aldrich)* were added to the phantom to control deoxygenation. Oxygen was injected through the inlet of the oxygen tank until the pO2 reached the maximum plateau, and measurements of the devices were performed under deoxygenation. The liquid phantom was stirred using a magnetic stir bar (200 rpm). To calculate $SO2_{brain}$, the liquid phantom was sampled using a syringe periodically. The $SO2_{brain}$ of the sample was derived from the pO2 and pH analyzed using a gas analyzer (i-smart 300, i-SENSE)[37]. The results of $SO2_{brain}$ were compared with those obtained from a commercial NIRS device and the proposed system in the range of $SO2_{brain}$ in human boundary conditions.

The O3 adult sensor was used as the commercial NIRS device. The proposed device was turned off while the commercial device was used to prevent interference between them, and vice versa. To minimize errors due to the timing of the measurement, $rSO2_{commercial}$ and $rSO2_{proposed}$ were obtained immediately after $SO2_{brain}$ sampling and before $SO2_{brain}$ sampling, respectively. The experiments were performed based on changes in the surface phantom (anatomy), total Hb concentration ($C_{HbT}$), and CSF depth, which refers to the distance between the brain-shaped film phantom and the surface phantom (Supplementary Table 5). All experiments were approved by the

Korean Red Cross in strict compliance with the Institutional Review Board (IRB) guidelines for research involving human blood (P01-201712-33-001-P4).

*Prospective observational study*

The observational studies on ischemic stroke patients and healthy subjects were approved by the Institutional Review Board of Seoul National University Hospital (B-2006-616-301, B-2101-658-302), and all participants provided their consent before the measurements were acquired. The proposed system was used to assess 37 healthy subjects (18 male, mean age = 52.3) who were asked to stay in supine positions. The measurements were performed on the left and right sides of their forehead region. Eight other patients (mean age = 72.2) were assessed for rSO2 using the proposed system while in a supine position. Ischemic stroke patients were recruited from the emergency room and stroke unit in the Seoul National University Bundang Hospital. The lesion side was measured before the recanalization treatment or within 24 hours of the treatment. Cleaning of the forehead region with an alcohol swab always preceded the acquisition of the measurements to prevent any disturbance due to foreign substances.

## Acknowledgements

This work was supported by National Research Foundation of Korea (NRF); Ministry of Science and Technology(2019R1A2C20860514); and R&D program of Korea Advanced Institute of Science and Technology(N1190225). This work made use of MRI dataset of OpenfMRI project. We acknowledge OBELAB Inc for use of frequency domain fNIRS device Imagent (ISS Inc.) and regional oximetry O3 (Masimo Inc.). The prospective observational study was made by use of facilities of Seoul National University Bundang Hospital.

## Author contributions

M.J, S.Y and H.B designed and developed the system. J.K and C.Y advised the indication and the performance of the developed system, led the prospective observational study, guided the analysis of the clinical data, and reviewed the manuscript. M.J and S.Y implemented the deep learning algorithm and monitoring software. M.J, S.Y, J.K and B.K developed the hardware. M.J, J.L, S.Y and B.K designed the biomimetic phantom and performed the experiments. H.B oversaw the project. M.J and H.B wrote and edited the manuscript.

## Additional information

Supplementary information is available for this paper. Additional materials for this paper can be requested from the authors

## Competing interests

C.Y is a member of the advisory board of YBrain Inc., ARPI Inc., and LG Electronics, Republic of Korea.

**Figure legends and figures**

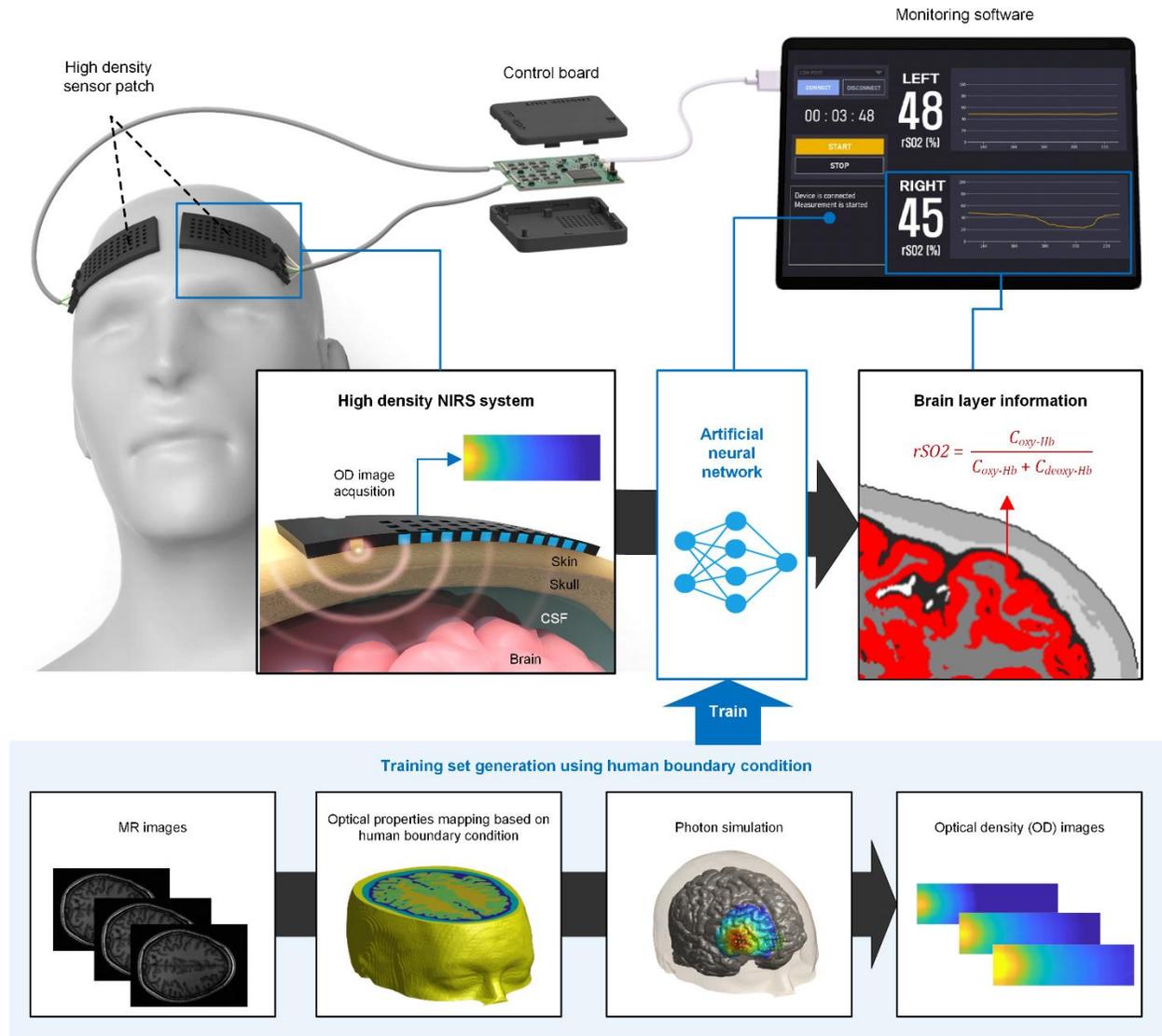

**Figure 1. Design of a patch-type brain layer analyzer, which includes a pair of flexible sensor patches, a control board, and monitoring software based on an artificial neural network.** The network extracts the brain layer information from the measured optical density images. The training set was generated using photon simulations performed on MR images.

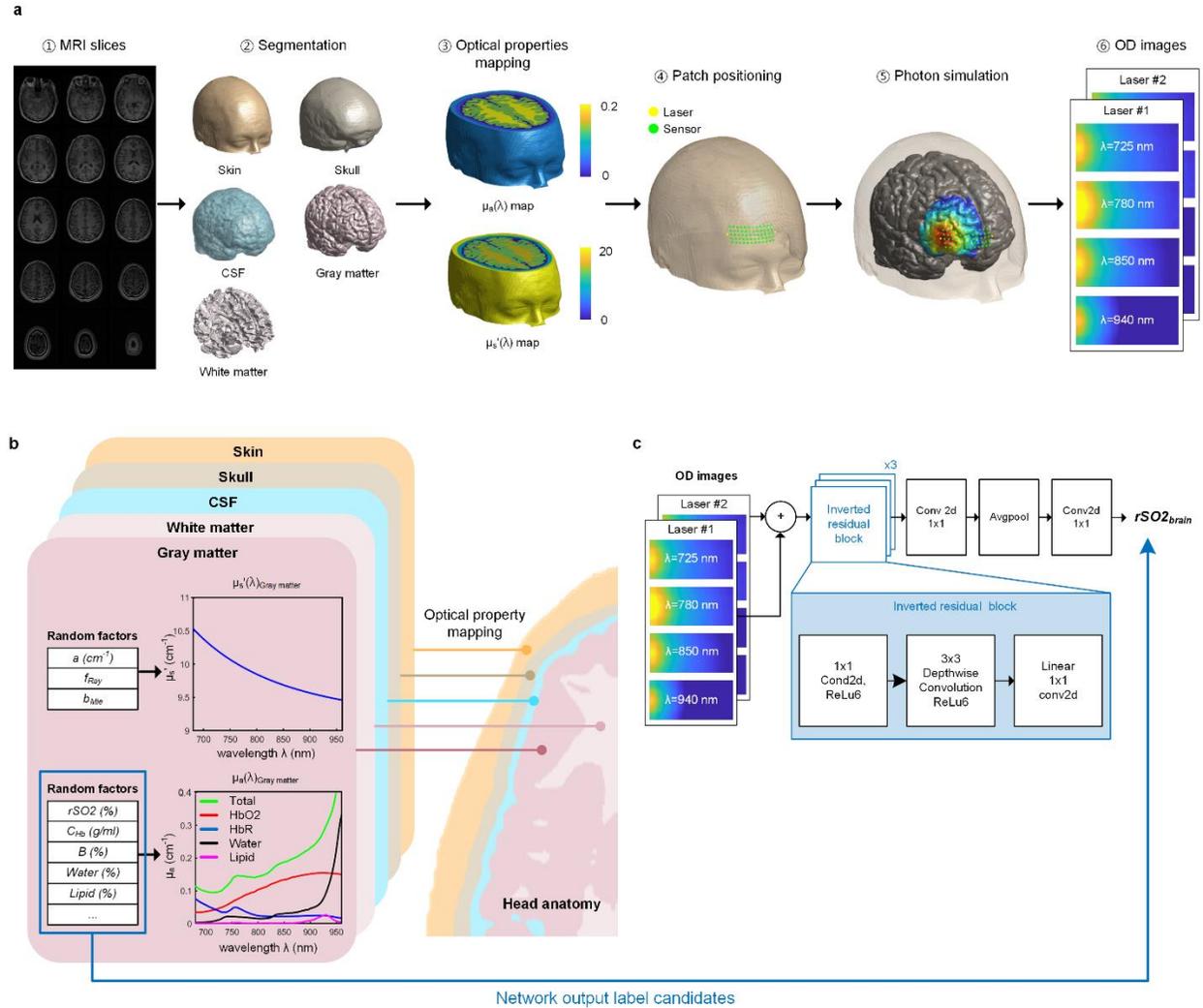

**Figure 2. Signal processing procedure of the brain layer analyzing network. a**, Dataset synthesis platform using MR slice and photon simulation. **b**, Mapping of optical properties into multilayer anatomy. **c**, Brain layer analyzing network based on the architecture of MobileNet V2.

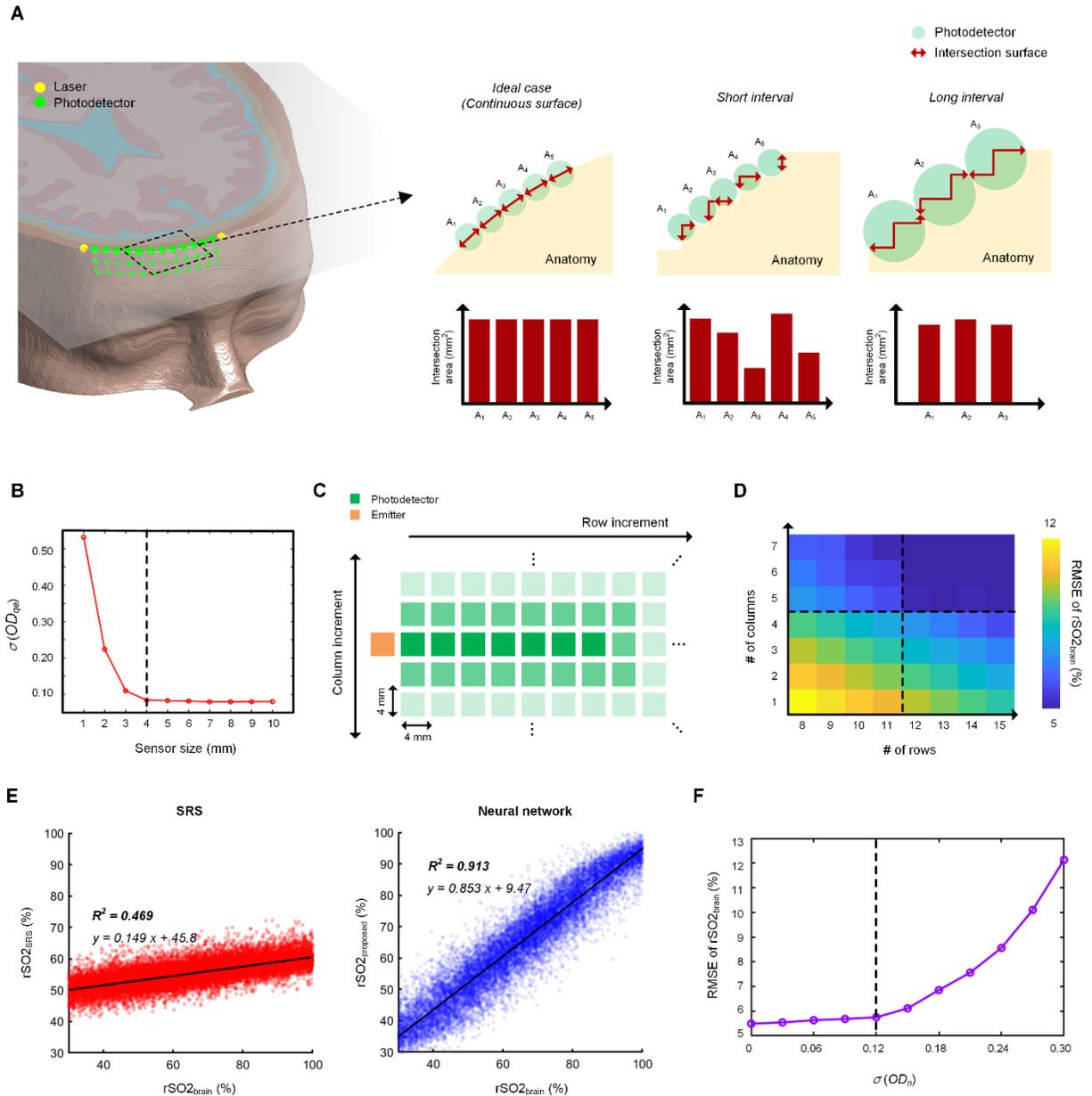

**Figure 3. Design strategy for the sensor array and performance of the brain analysis network.**

**a,** Quantization variance between photodetectors in photon simulation. **b,** The quantization errors of *OD* from the photodetectors with respect to the sensor interval. **c,** Changes in the numbers of rows and columns of the sensor array to investigate the optical arrangement. **d,** RMSE for rSO2$_{brain}$ with respect to the number of rows and columns of the sensor array. **e,** rSO2 extraction performance of the SRS technique and the proposed neural network. **f,** The performance of the network under noisy OD conditions.

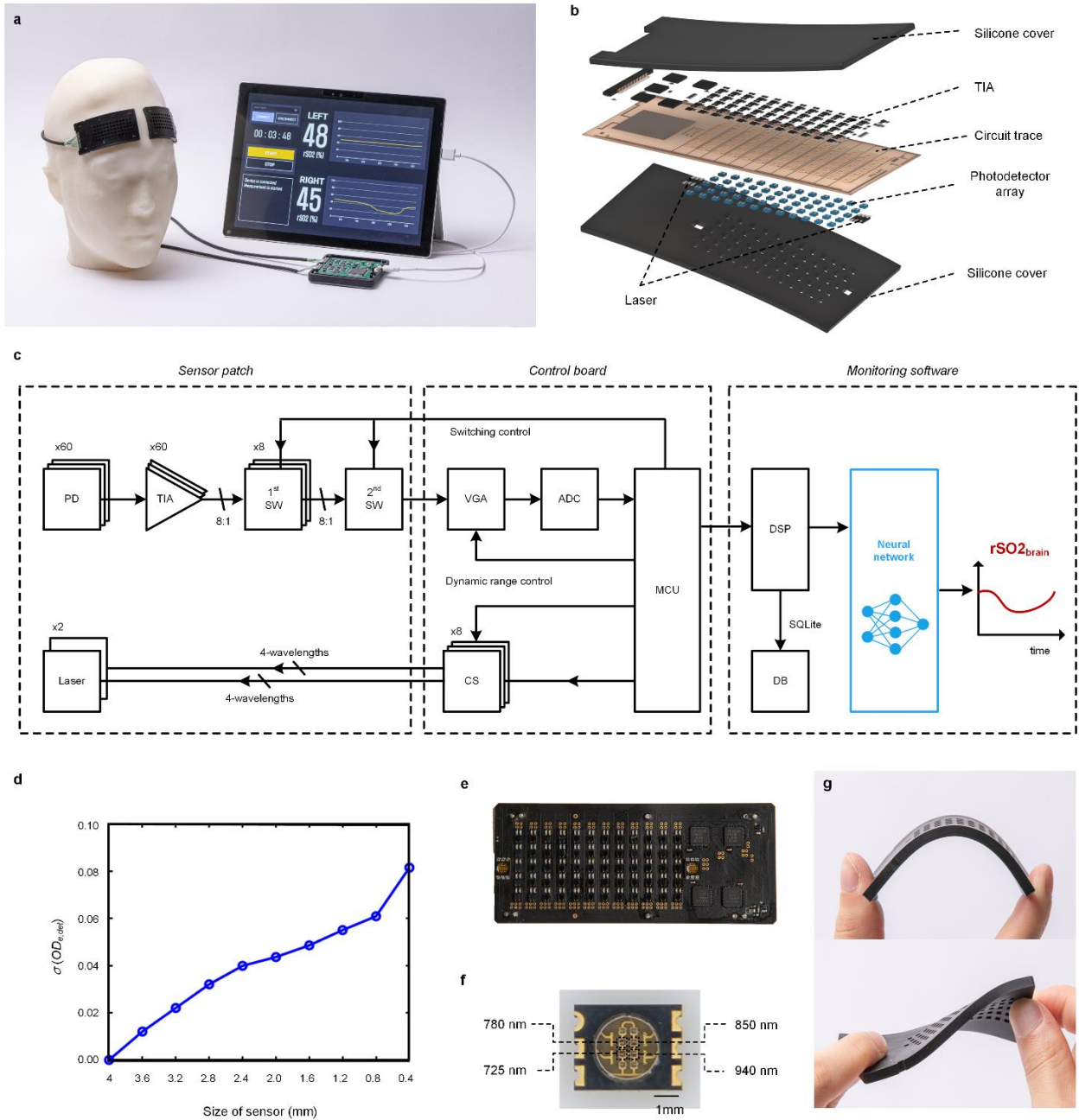

**Figure 4. High-density NIRS system for continuous monitoring of rSO2 in the cortex. a,** Proposed system consisting of sensor patches, a control board, and monitoring software, **b,** Expanded-view illustration of the constituent layer: silicone encapsulation layer, electronics, and flexible circuit trace. The electronics include the sensor array, laser diode, and transimpedance amplifier (TIA). **c,** System-level block diagram of sensor patches, a control board, and monitoring

software. **d,** $\Delta$**OD** of the shortest (4mm) channel with change of the sensor size. **e,** Photograph of the flexible patch. **f,** Photograph of custom designed four-wavelength laser diodes (725, 780, 850 and 940 nm). **g,** Photographs of the patch upon mechanical stresses of bending and twisting.

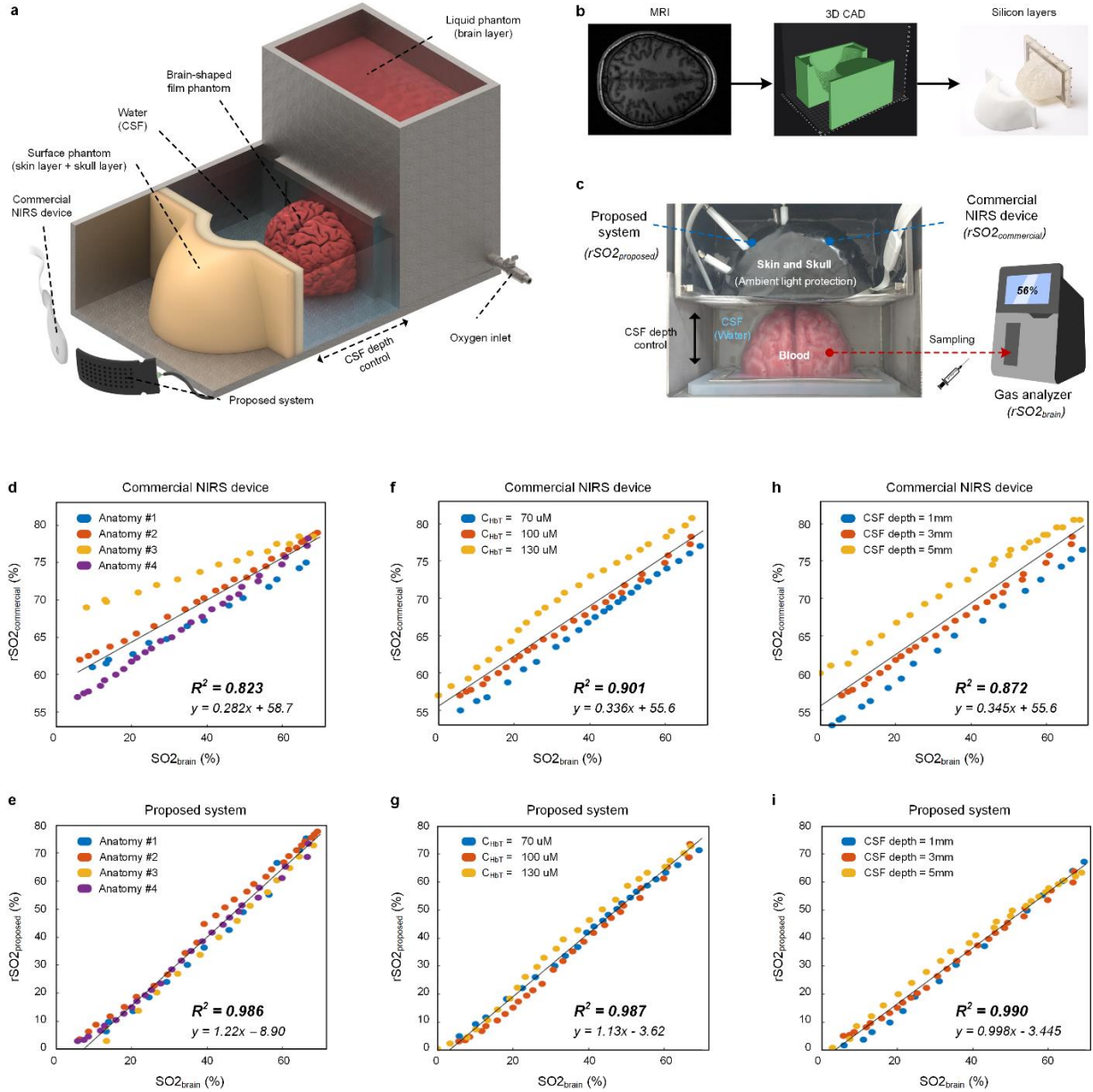

**Figure 5. Experimental studies on biomimetic phantoms. a,** Multilayer brain phantom consists of surface phantom, water, brain-shaped film phantom, and liquid phantom. The surface phantom represents the skin and skull layers with different optical properties. The liquid phantom and water represent the brain and CSF layer, respectively, while the transparent film layer with a thickness of 1 mm separates the two fluid layers. The proposed patch and the commercial NIRS device are attached to the surface phantom. **b,** Fabrication process of the surface phantom and the film

phantom using MR images. **c,** Photograph of the experimental setup used to extract rSO2 from the brain cortex under structural variations. **d-i,** Measured rSO2 of the proposed system and the commercial NIRS device under different surface phantoms (**d, e**), different blood volumes in the brain layer (**f, g**), and different thicknesses of the CSF layer (**h, i**).

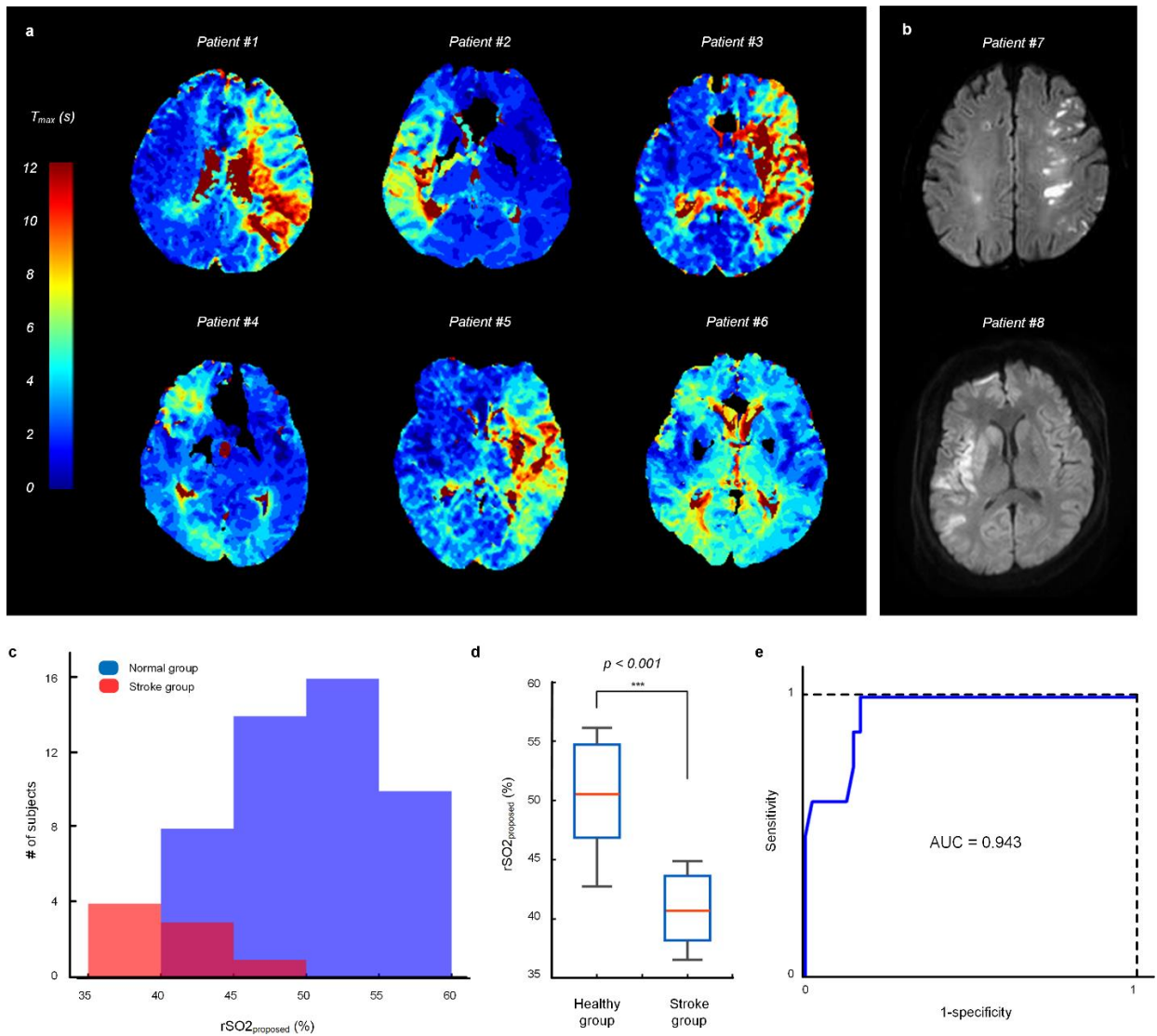

**Figure 6.** Prospective observational studies of rSO2 in healthy subjects and ischemic stroke patients. **a-b**, Perfusion MR imaging maps and diffusion-weighted images of patients. **c**, Histogram of the rSO2 in both groups. **d**, Box plot of the rSO2 in both groups and their significant difference (paired t-test). **e**, Receiver operating characteristic curve used to distinguish between healthy subjects and stroke patients.

# AI-Enhanced High-Density NIRS Patch for Real-Time Brain Layer Oxygenation Monitoring in Neurological Emergencies

Minsu Ji[1], Jihoon Kang[2], Seongkwon Yu[1], Jaemyoung Kim[3], Bumjun Koh[1], Jimin Lee[1], Guil Jeong[1], Jongkwan choi[3], Chang-Ho Yun[2†], and Hyeonmin Bae[1†]*

[1]School of Electrical Engineering, Korea Advanced Institute of Science and Technology, Daejeon, Republic of Korea

[2]Department of Neurology, Seoul National University Bundang Hospital and Seoul National University College of Medicine, Seongnam, Gyeonggi-do, Republic of Korea

[3]Department of Research and Development, Optics Brain Electronics Laboratory, OBELAB Inc, Seoul, Republic of Korea

[†]The two authors share the senior authorship.

**\*Corresponding author**

E-mail: hmbae@kaist.ac.kr

**Supplementary Information**

Supplementary Note 1-6

Supplementary Figures S1-S12

Supplementary Tables S1-S5

Reference

**Supplementary Note 1**

The simulated light intensity $\hat{I}_{sim}$ of the photodetector is determined by

$$\hat{I}_{sim} = E_e \hat{A} = E_e(A + A_{qe}) = E_e A (1 + r_{qe}) = I_{sim}(1 + r_{qe})$$

where $E_e$ is the irradiance flux density (W/m²), $\hat{A}$ is the intersection area between the photodetector and the surface of the MR image, $A$ is the ideal intersection area of the photodetector, $A_{qe}$ is the quantization error in the intersection area of the photodetector, $I_{sim}$ is the error-free underlying light, and $r_{qe}$ is the quantization error ratio, $A_{qe}/A$. Supplementary figure 1 shows the quantization error resulting from the resolution limitation of the ground truth MR image.

The variance in the intersection areas among the photodetectors finally results in the quantization error in the simulated $OD$, which can be expressed by

$$\widehat{OD} = OD + OD_{qe} = -ln(I_{sim}) - ln(1 + r_{qe})$$

Under the condition of steady state and $r_{qe} \ll 1$, the standard deviation in $OD$ can be estimated by

$$\sigma(OD_{qe}) = \sigma\left(ln(1 + r_{qe})\right) \cong \sigma(r_{qe})$$

The $\sigma(r_{qe})$ was found by sensor array positioning on the real MR images of 600 subjects. The finite element model (FEM) was used to calculate the intersection area (Supplementary Figure 1). Figure 3b represents the variance in the $OD_{qe}$ obtained from the variance in the $r_{qe}$ while varying the size of the photodetector. The chosen dimension of each photodetector is 4 mm, which is sufficiently large to be desensitized by the pixelation of the MR image.

**Supplementary Note 2**

To compare the performances of the proposed method and the conventional SRS technique, the calculation procedure used for the SRS should be considered. The optical properties of the measured target were determined from the multi-distance measurement of $OD(L)$. In particular, the $\mu_{eff}$ of the measured subject is derived by differentiating $OD(L)$ with respect to L, as represented by

$$OD(L) = \mu_{eff}L + 2\ln(L) + C$$

$$\frac{\partial(OD(L) - 2\ln(L))}{\partial L} = \mu_{eff} \cong \sqrt{3\mu_a\mu_s'}$$

where OD is the optical density, L is the distance between the source and detector, $C$ is the constant derived from the photon diffusion equation, and $\mu_{eff}$ is the effective attenuation coefficient defined as $\sqrt{3\mu_a(\mu_s' + \mu_a)}$[1-3]. It can be approximated as $\sqrt{3\mu_a\mu_s'}$, since $\mu_s' \gg \mu_a$ in the high scattering medium. As long-distance channels are more sensitive to the optical properties of the brain tissue, we used the channels with L > 3 cm to calculate the slope of OD, which is the method adapted by commercially available SRS devices. Supplementary Figure 12 illustrates the process of determining $\mu_{eff}$ using polynomial fitting. After performing polynomial fitting to calculate $\mu_{eff}(\lambda)$ for all wavelengths, the subsequent steps are used to convert the results into rSO2.

The $\mu_s'(\lambda)$ value can be assumed to be $k(1 - h\lambda)$ in the near infrared region, where k is the unknown constant factor; therefore, $k\mu_a(\lambda)$ can be represented by

$$k\mu_a(\lambda) = \frac{1}{3(1 - h\lambda)}\mu_{eff}(\lambda)^2$$

The value of $h$ used in this study was 6.3x10$^{-4}$ (mm-1/nm). The relative concentrations and fraction of biological tissue can be calculated by

$$\begin{bmatrix} kC_{HbO_2} \\ kC_{HbR} \\ kF_{H_2O} \end{bmatrix} = [\varepsilon]^{-1} \begin{bmatrix} k\mu_a(\lambda_1) \\ k\mu_a(\lambda_2) \\ k\mu_a(\lambda_3) \\ k\mu_a(\lambda_4) \end{bmatrix}$$

where $C_{HbO_2}$ is the concentration of oxyhemoglobin, $C_{HbR}$ is the concentration of deoxyhemoglobin, $F_{H_2O}$ is the volume fraction of the water, and $[\varepsilon]$ is the matrix of extinction coefficients. Wavelengths of 725, 780, 850, and 940 nm were used to calculate the number of chromophores. The extinction coefficients were retrieved from previous works. The rSO2 is derived from the relative concentrations, as shown in the following equation.

$$rSO_2 = \frac{kC_{HbO_2}}{kC_{HbO_2} + kC_{HbR}} = \frac{C_{HbO_2}}{C_{HbO_2} + C_{HbR}}$$

**Supplementary Note 3**

The error of the measured light can be modeled as

$$\hat{I} = I(1 + r_e)$$

where $\hat{I}$ is the measured light, $I$ is the error-free underlying light, and $r_e$ is the ratio of error. The noise corrupted OD can be represented by

$$\widehat{OD} = OD + OD_e = -log(I) - log(1 + r_e)$$

where $OD_e$ is the measurement error in $OD$. Under the condition of steady state and $r_e \ll 1$, the standard deviation in OD can be expressed by

$$\sigma(OD_e) = \sigma(log(1 + r_e)) \cong \sigma(r_e)$$

Given $\sigma(OD_e)$, the SNR can be calculated by the following equation.

$$SNR = 20log\left(\frac{E(I)}{\sigma(Ir_e)}\right) = 20log\left(\frac{1}{\sigma(r_e)}\right) = 20log\left(\frac{1}{\sigma(OD_e)}\right)$$

As shown in Figure 3f, the brain analyzing network demonstrated stable performance with a $\sigma(OD_e)$ of 0.12.

The $OD_e$ can be estimated by the following equation:

$$OD_e = OD_{noise} + OD_{e,det}$$

where $OD_{noise}$ represents the electrical white noise and $OD_{e,det}$ is the error due to a reduction in detector size. Although the detector pixel size is set at 4mm in the brain layer analyzing network to mitigate the quantization error in the simulation, we have decreased the actual sensor size to 2mm for the sake of hardware flexibility. This reduction has introduced an additional error of $\sigma(OD_{e,det})$ of 0.04 compared to the ideal simulation result, as illustrated in Figure 4c. As a result, the allowed electrical noise $\sigma(OD_{noise})$ is 0.113, which is derived from

$\sqrt{\sigma(OD_e)^2 - \sigma(OD_{e,det})^2}$. Finally, we set the target SNR is set at 18.9 dB, which is calculated from $20log\left(\frac{1}{0.113}\right)$.

**Supplementary Note 4**

The 600,000 distinctive data from the dataset synthesis platform were used to estimate the required dynamic range of the system. Supplementary figure 2 represents the histogram of the light attenuations of the photodetectors respect to the shortest channel (4mm) for all data. The dynamic range of 140 dB is chosen, which covers entire dataset.

The high density NIRS system achieves a 140 dB dynamic range through the use of the VGA, ADC, and dynamic laser control (Supplementary Fig. 4). The total dynamic range $DR_{total}$ can be expressed by

$$DR_{total} = DR_{laser} + DR_{VGA} + DR_{ADC,eff}$$

where $DR_{laser}$ and $DR_{VGA}$ are the dynamic ranges of laser emitting system and VGA, and $DR_{ADC,eff}$ is the effective dynamic range of ADC, considering the 11.8 dB SNR (Supplementary Note 6). The $DR_{VGA}$ is 50 dB, and $DR_{ADC,eff}$ can be calculated as

$$DR_{ADC,eff} = 20\log\left(\frac{1.65\ V}{4.04\ mV}\right) = 52\ dB$$

where 1.65 V is the full-scale voltage, and 4.04 mV is the noise level at the minimum photocurrent (Supplementary Figure 5). In order to achieve the 140 dB dynamic range, $DR_{laser}$ must be over 38 dB. With a maximum average output power of 5 mW, which satisfies the human safety standard (Laser class 3R), the laser emitting system controls the output power from 60 µW to 5mW, resulting in a $DR_{laser}$ of 38 dB. The limited output range of the stimulated emission region of laser diode is also supplemented by utilizing the spontaneous emission region of the laser diode is also supplemented by utilizing the spontaneous emission region to cover small output powers (Supplementary figure 5).

**Supplementary Note 5**

To construct the 2D OD image, the relative ratio among the entire sensors should be calculated. Supplementary figure 4.b shows the LD current and VGA gain switching stages after the PD switching to cover the 140 dB dynamic range. However, effective gain of each stage cannot be calculated due to the non-linearity of spontaneous emission and sample variances in the laser diodes (see Supplementary Figure 5 for details). To compensate this non-linearity, the gains of each stage were obtained by the measurement results.

The gain of each stage normalized by the gain of *stage 1* can be expressed by

$$G_{stage\ 1} = 1$$

$$G_{stage\ 2} = G_{stage\ 1} \times \frac{1}{n_{1,2}} \sum_{i}^{n_{1,2}} \frac{I_{i,stage\ 2}}{I_{i,stage\ 1}}$$

$$G_{stage\ 3} = G_{stage\ 2} \times \frac{1}{n_{2,3}} \sum_{i}^{n_{2,3}} \frac{I_{i,stage\ 3}}{I_{i,stage\ 2}}$$

$$\ldots$$

$$G_{stage\ N} = G_{stage\ N-1} \times \frac{1}{n_{N-1,N}} \sum_{i}^{n_{N-1,N}} \frac{I_{i,stage\ N}}{I_{i,stage\ N-1}}$$

where $G_{stage\ N}$ is the gain of *stage N* normalized by the gain of *stage 1*, $n_{N-1,N}$ is the number of measurable detectors at the both of *stage N-1* and *stage N* (Orange region in the Supplementary Figure 4.c), $I_{i,stage\ N}$ is the measured intensity on the $i^{th}$ detector at the *stage N*.

The effective intensity $I_{eff}$ considering the gain of each stage can be denoted by

$$I_{i,eff,} = G_{stage\ N}\ I_{i,stage\ N}$$

The OD image can be expressed by

$$\vec{OD} = -\ln(\frac{\vec{I_{eff}}}{I_0})$$

where $I_0$ is the incident light. Since incident light $I_0$ is unknown due to the ambiguity in the direction of the light and the reflective coefficient of skin, The OD image was normalized by the intensity $I_{1,eff}$ at the shortest channel (4mm). The final normalized OD image can be represented by

$$\vec{OD}_{norm} = -\ln(\frac{\vec{I_{eff}}}{I_{1,eff}})$$

**Supplementary Note 6**

To ensure robust performances of the brain analyzing network, the system should achieve the SNR of 18 dB even in the worst-case subject with the -140dB attenuation in respect to the shortest 4 mm channel (Supplementary Note 4). The experimental setup was built to investigate the photocurrent on the worst-case anatomy. The commercial optical phantom (Biomimic, INO) was employed to provide the -140 dB light attenuation. The power of laser diode was controlled under 5 mW satisfying the laser class 3R. The photocurrent measurement result was 5 pA.

The noise analysis on the designed high-density NIRS system was performed to estimate the SNR of the system for the worst-case (Supplementary Fig. 5). The estimated SNR after the ADC stage was 11.8 dB due to the thermal noise on the TIA which is the dominant noise source in the overall system (TIA thermal noise: 405 nV/Hz$^{0.5}$, OP-amp input referred voltage noise: 27 nV/Hz$^{0.5}$, VGA voltage noise: 5 nV/Hz$^{0.5}$). The digital oversampling was employed to suppress the noise, achieving the final SNR of 20.8dB over the target SNR of 18.9 dB.

The noise of the implemented system was measured under the experimental setup with the -140 dB attenuation. The measured SNR of the system was 20.1 dB which is in agreement with the noise analysis (Supplementary Fig. 6).

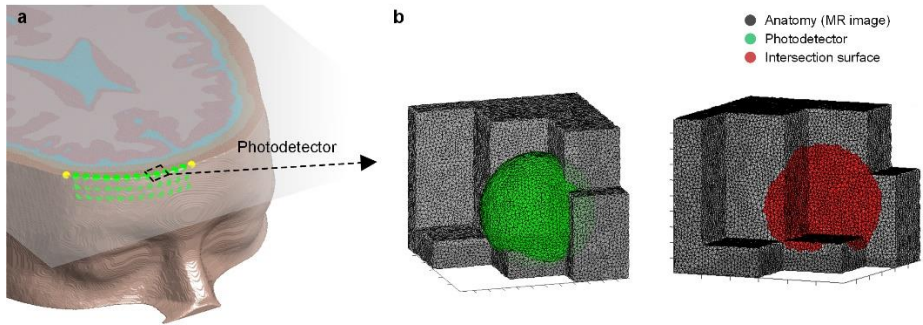

**Supplementary Figure 1. a**, Sensor array positioning on the anatomy based on MR images. **b**, FEM analysis to estimate intersection area between the simulated photodetector and MR image.

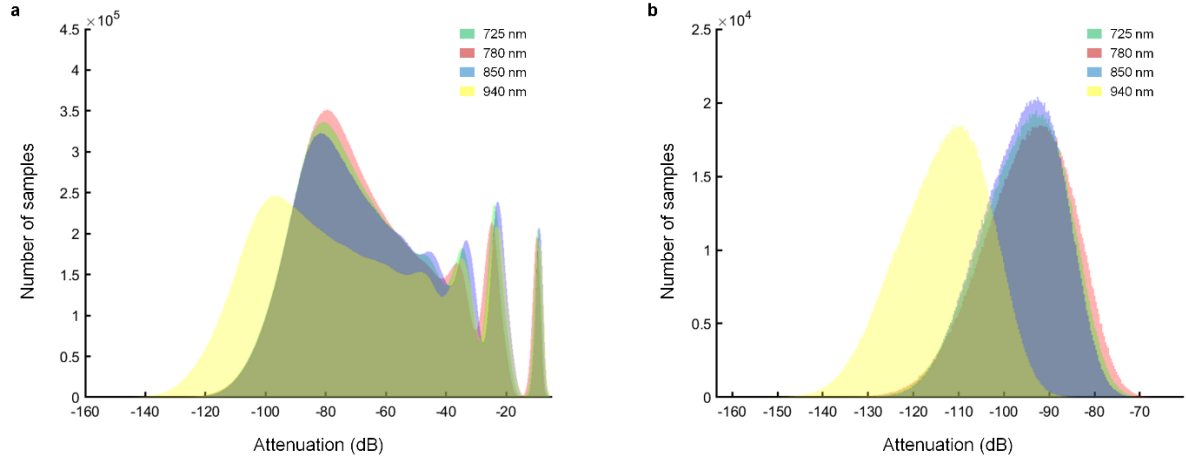

**Supplementary Figure 2. a**, Histogram of the light attenuation of all photodetectors with respect to the shortest channel (4mm) for all dataset from dataset synthesis platform. **b**, Histogram of the light attenuation of the longest channels (48.6 mm) with respect to the shortest channel (4mm) for all dataset.

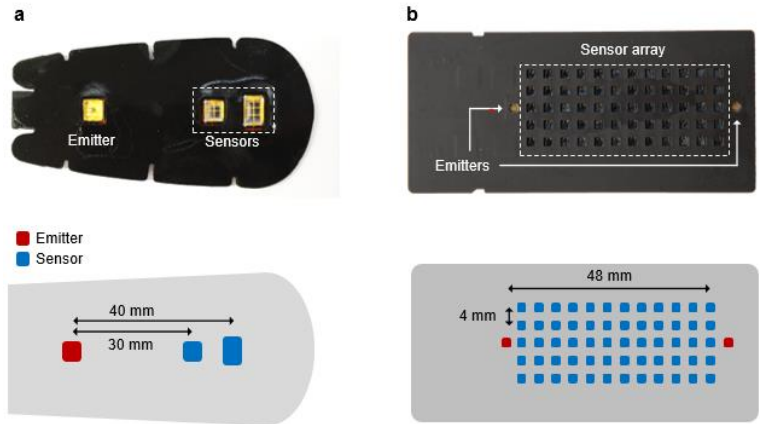

**Supplementary Figure 3. a**, Sensor array of the commercial NIRS device (O3, Masimo). **b**, Sensor arrangement of the proposed high-density patch.

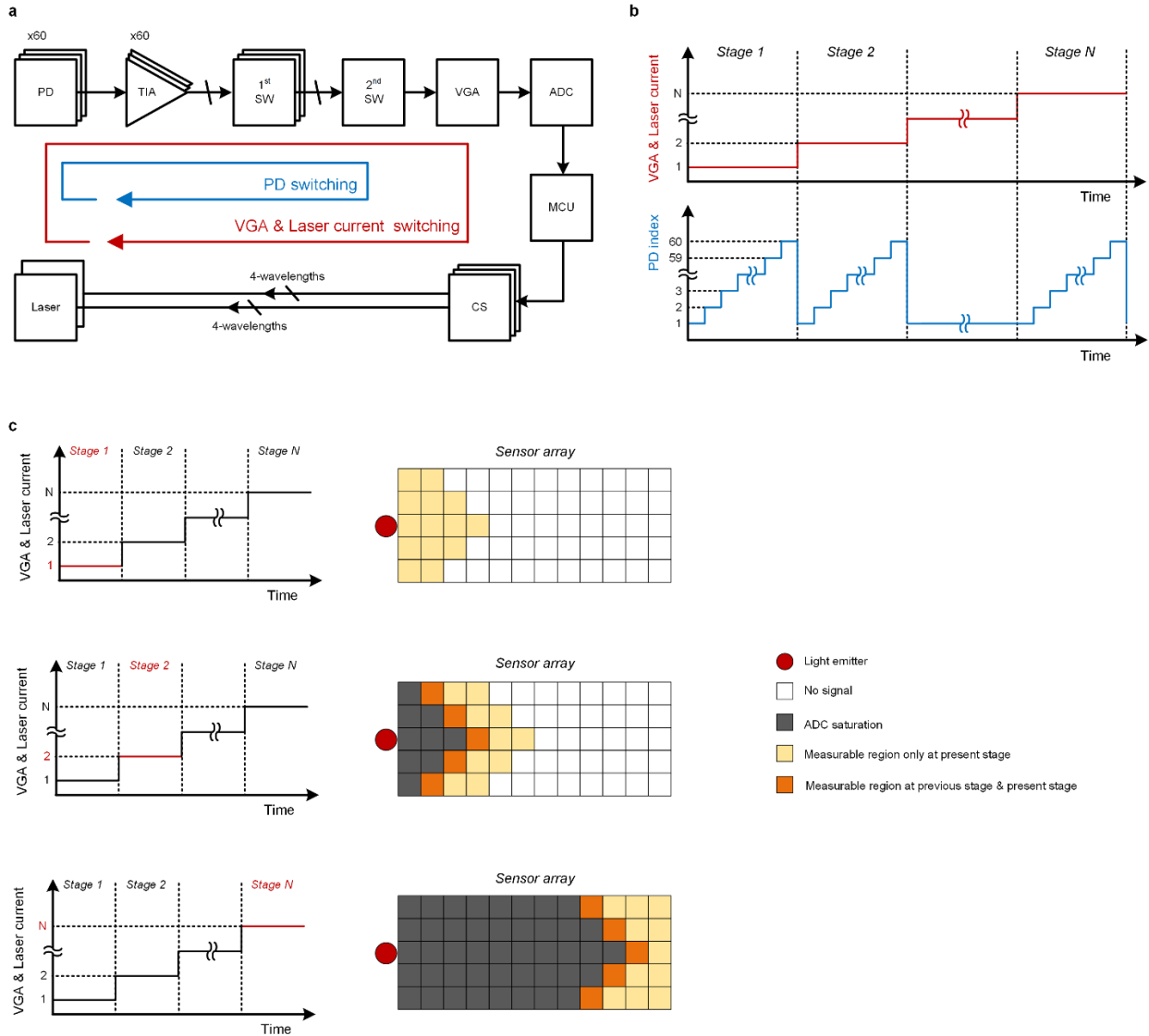

**Supplementary Figure 4. a**, Block diagram of the system with the laser power and VGA gain switching after the PD switching. **b**, Gain stages with the increment in laser power and VGA gain. **c**, the measurable regions in the sensor array with the gain stages.

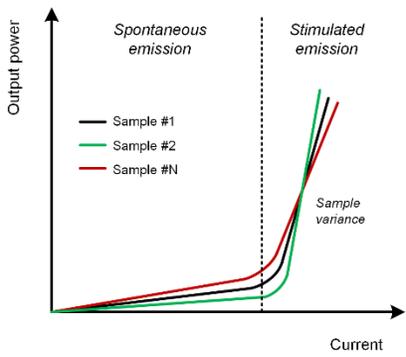

**Supplementary Figure 5.** Laser output power vs. Current.

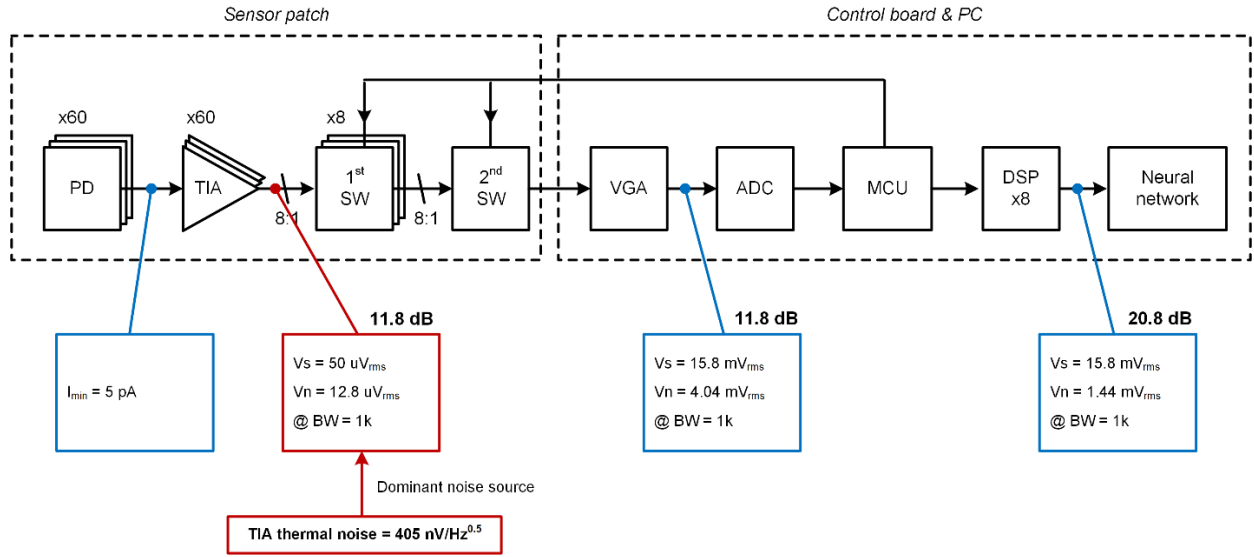

**Supplementary Figure 6.** Noise analysis of the entire system for the minimum photocurrent case.

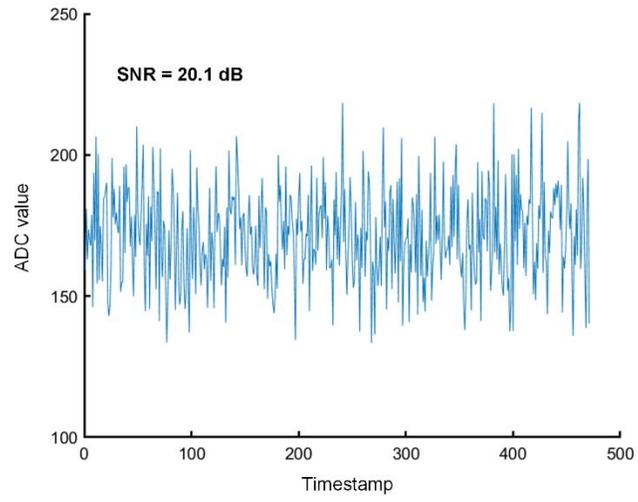

**Supplementary Figure 7.** SNR of the measurement result from High-density NIRS system.

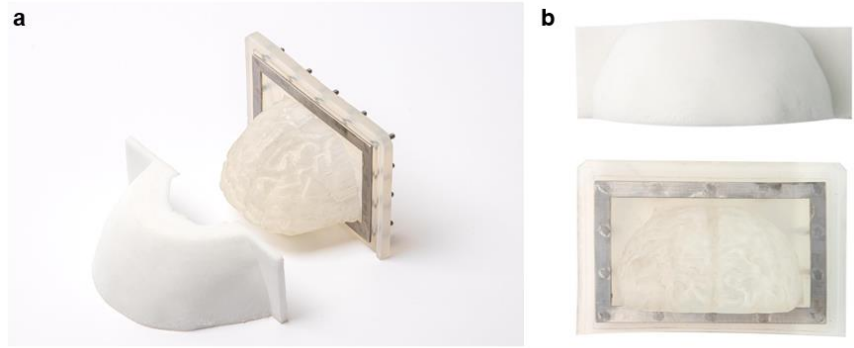

**Supplementary Figure 8. a**, Diagonal views of the surface phantom and brain-shaped film phantom. **b**, Front views of the surface phantom and brain-shaped film phantom

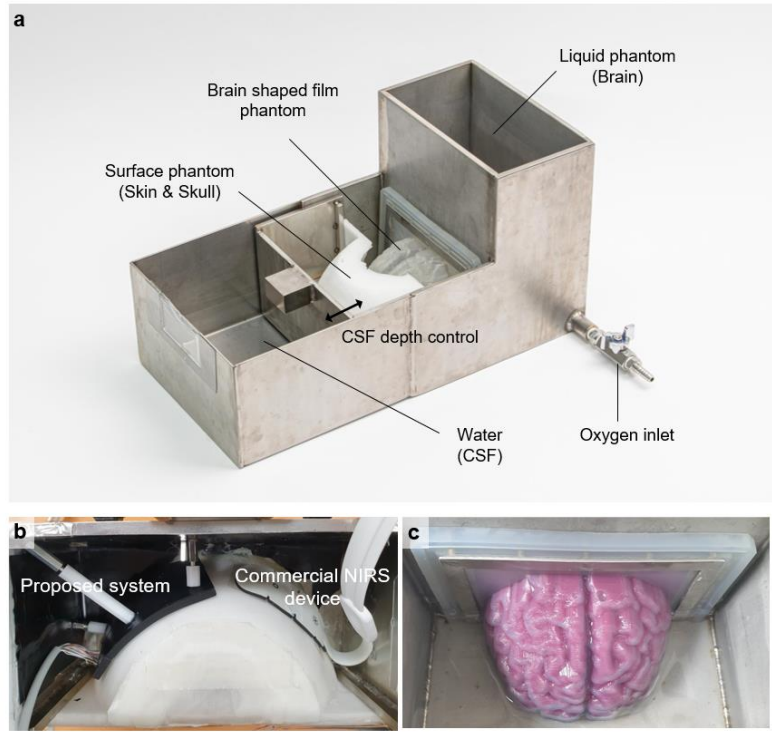

**Supplementary Figure 9. a**, Experimental setup for the multilayer phantom. **b**, Attachments of the proposed patch and commercial NIRS device (O3, Masimo) on the surface phantom. **c**, Brain-shaped film phantom with the blood phantom in the brain layer.

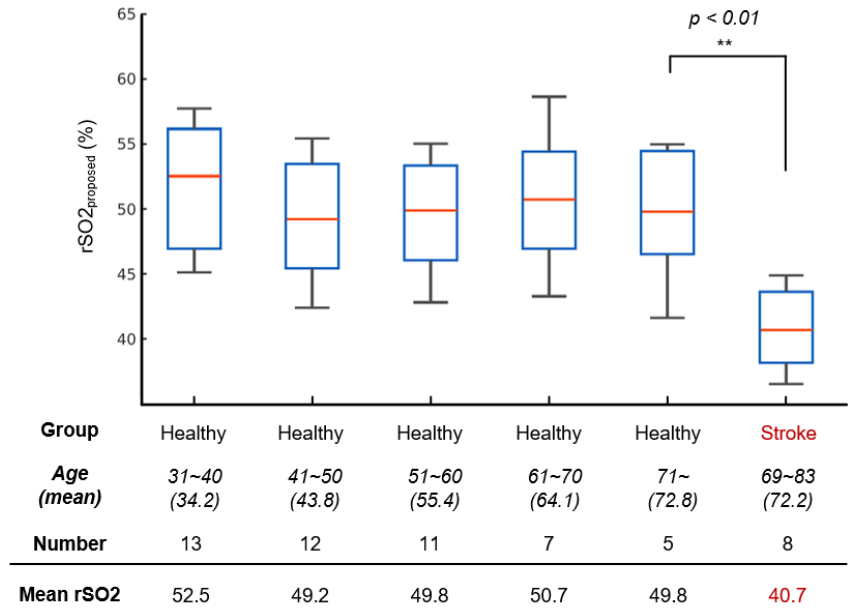

**Supplementary Figure 10.** The rSO2 results with respect to age in the healthy subjects. The red line indicates the median of the boxplot, whereas the blue box indicates the 25th and 75th percentiles of the results.

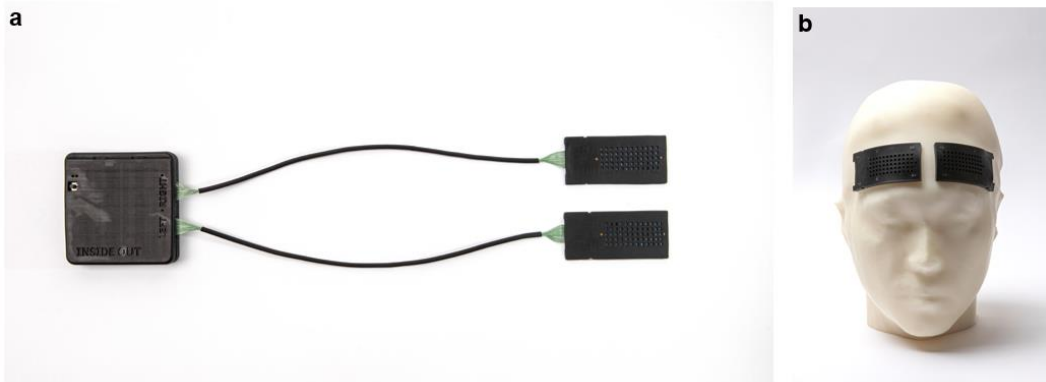

**Supplementary Figure 11. a**, Implementation of the system consisting of a pair of patches and the control board. **b**, Attachment of patches on the forehead region.

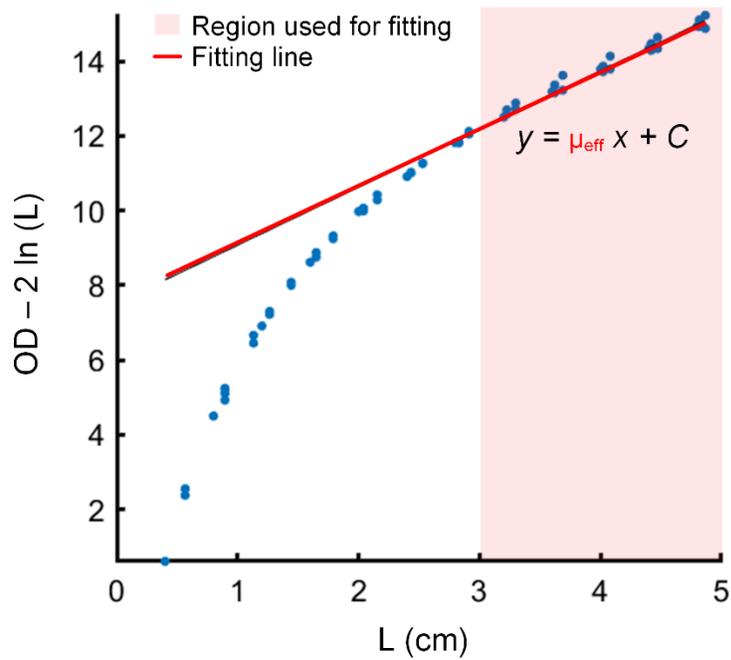

**Supplementary Figure 12.** Polynomial fitting method for calculating $\mu_{eff}$ in SRS. To calculate $\mu_{eff}$, a linear fitting was performed between the channel length $L$ and $OD(L) - 2\ln(L)$. The slope of this linear fit was approximated as $\mu_{eff}$. To minimize surface-related errors, only channels with lengths greater than 3 cm were used in the fitting process.

**Supplementary Table 1.** Boundary condition of biological components for the absorption coefficient.

| Factor | Skin | Skull | CSF | Gray matter | White matter |
|---|---|---|---|---|---|
| $C_{Hb,blood}$ (mM) | 1.68-2.62 | | | | |
| B (%) | 0.1-0.4 | 0.5-1.8 | 0 | 1-5 | 1-5 |
| rSO2 (%) | 40-100 | 50-80 | 0 | 0-80 | 0-80 |
| W (%) | 20-50 | 5-20 | 100 | 65-90 | 65-90 |
| F (%) | 10-30 | 30-50 | 0 | 3-12 | 5-24 |
| M (uM) | 0-200 | 0 | 0 | 0 | 0 |
| N (%) | 0 | 10-25 | 0 | 0 | 0 |

**Supplementary Table 2.** Boundary condition of the scattering coefficient

| Factor | Skin | Skull | CSF | Gray matter | White matter |
|---|---|---|---|---|---|
| $a$ | 36-58 | 9.7-20.9 | 0.01 | 13.3-15.7 | 28-34 |
| $f_{Ray}$ | 0.22-0.7 | 0-0.04 | 0 | 0.36-0.53 | 0.74-0.9 |
| $b_{Mie}$ | 0.91-1.2 | 0.12-0.54 | 0 | 0 | 0 |

**Supplementary Table 3.** Comparison of the proposed high-density patch with commercial NIRS devices

|  | Invos (Medtronic) | O3 (Masimo) | Proposed |
|---|---|---|---|
| Number of emitters | 1 | 1 | 2 |
| Number of wavelengths | 2 | 2 | 4 |
| Number of detectors (N) | 2 | 2 | 60 |
| Number of channels | 4 | 4 | 480 |
| Measurement area (cm$^2$) | 8 | 8 | 20.8 |
| Channel density (N/cm$^2$) | 0.5 | 0.5 | 23.07 |

**Supplementary Table 4.** Optical properties of surface phantoms

| Surface phantom | | #1 | #2 | #3 | #4 |
|---|---|---|---|---|---|
| Skin layer | India Ink (uL/100 ml) | 3 | 5 | 15 | 30 |
| | TiO2 (g/100 ml) | 0.4 | 0.4 | 0.4 | 0.4 |
| | $\mu_{eff}(725nm)$ | 1.04 | 1.02 | 1.29 | 1.72 |
| | $\mu_{eff}(780nm)$ | 0.99 | 0.95 | 1.25 | 1.69 |
| | $\mu_{eff}(850nm)$ | 1.00 | 0.95 | 1.24 | 1.70 |
| | $\mu_{eff}(940nm)$ | 1.07 | 1.04 | 1.28 | 1.74 |
| Skull layer | India Ink (uL/100 ml) | 0 | 10 | 15 | 20 |
| | TiO2 (g/100 ml) | 0.4 | 0.4 | 0.4 | 0.4 |
| | $\mu_{eff}(725nm)$ | 0.84 | 1.08 | 1.23 | 1.49 |
| | $\mu_{eff}(780nm)$ | 0.81 | 1.03 | 1.16 | 1.44 |
| | $\mu_{eff}(850nm)$ | 0.81 | 1.02 | 1.17 | 1.46 |
| | $\mu_{eff}(940nm)$ | 0.87 | 1.1 | 1.24 | 1.49 |

**Supplementary Table 5.** Experimental conditions for the multilayer phantom experiments

| Experiment # | Surface phantom | $C_{HbT}$ (uM) | CSF depth (mm) |
|---|---|---|---|
| 1-1 | #1 | 100 | 3 |
| 1-2 | #2 | 100 | 3 |
| 1-3 | #3 | 100 | 3 |
| 1-4 | #4 | 100 | 3 |
| 2-1 | #4 | 70 | 3 |
| 2-2 | #4 | 100 | 3 |
| 2-3 | #4 | 130 | 3 |
| 3-1 | #4 | 100 | 1 |
| 3-2 | #4 | 100 | 3 |
| 3-3 | #4 | 100 | 5 |